\documentclass[prb,amsmath,amssymb,showpacs,twocolumn]{revtex4}

\usepackage{graphicx}
\def\be{\begin{equation}}
\def\ee{\end{equation}}
\def\c{{c\uparrow}}

\begin{document}
\title{Spontaneous superconducting islands and Hall voltage in clean superconductors}
\author{Jorge Berger}
\affiliation{Physics Unit, Ort Braude College, P. O. Box 78,
21982 Karmiel, Israel and \\
Department of Physics, Technion, 32000 Haifa, Israel}
\email{phr76jb@tx.technion.ac.il}
\begin{abstract}
We study a clean superconductor in the Hall configuration, in the 
framework of a purely dissipative time-dependent Ginzburg--Landau model. 
We 
find situations in which the order parameter differs significantly from zero in a set of islands that appear to form a periodic structure. When the pattern of islands becomes irregular, it moves in or against the direction of the current and a Hall voltage is found. Tiny differences in the initial state may reverse the sign of the Hall voltage. When the average Hall voltage vanishes, the local Hall voltage does not necessarily vanish. We examine the influence that several boundary conditions at the electrodes have on these effects.
\end{abstract}
\pacs{74.20.De, 74.25.Qt, 74.25.Op, 74.78.Db}%
\maketitle
\section{INTRODUCTION}
The Hall voltage in superconductors exhibits a rich variety of behaviors.\cite{Kop}
In the Meissner state there is no Hall voltage,\cite{Lewis} but in the mixed state Hall voltage is present due to vortex drag.\cite{th,exp} In some cases, the sign of the Hall voltage is opposed to what would naively be expected.\cite{anom}

We shall consider a thin rectangular superconducting sample. Let a magnetic field be applied in the $z$-direction and let a total current $I$ flow in the $x$-direction. The sample will be assumed to be sufficiently long in the $z$-direction, so that physical quantities will be independent of $z$. The current will be assumed to flow in the entire range $-\infty<x<\infty$, but only the segment $0\le x\le L$ will be superconducting. We denote the thickness of the sample by $d$; the regions $y<0$ and $y>d$ are taken as insulating.
We will study the current dependence of measurable quantities within the 
framework of the time-dependent Ginzburg-Landau model (TDGL).\cite{TDGL} 
Among the diversity of formulations of TDGL, we consider the simplest: \be
{\partial_t \psi }=-\frac 1\eta \left[ \left( -i\nabla -%
{\bf A}\right) ^2\psi +\left( 1-T\right) \left( |\psi |^2-1\right) \psi
\right] +\tilde{f} \;, \label{psi} 
\ee
\be
{\partial_t {\bf A}}=\left( 1-T\right) 
{\rm Re} \left[ {\bar\psi}
\left( -i\nabla -{\bf A}\right) \psi \right] -\kappa ^2\nabla \times
\nabla \times {\bf A}\;.  \label{ohm}
\ee
Here $\psi$ is the order parameter, $t$ is the time, ${\bf A}$ is the electromagnetic vector potential, $\eta $ is the ratio between the relaxation times of $\psi $ and ${\bf A}$, $\kappa $ is the Ginzburg-Landau parameter, $T$ is the temperature and $\tilde{f}$ a random ``force" that simulates thermal fluctuations. The units are customary, as e.g. in Refs.~\onlinecite{Kato,Bolech,CS,Fortin}. The gauge is chosen such that the scalar potential is zero.

We shall see that this configuration leads to the appearance of a phenomenon which, to my knowledge, has not been previously encountered: spots where $|\psi|$ is significant, whereas most of the sample is practically in the normal state. We call these spots ``superconducting islands." The expression ``superconducting islands"  is usually intended for regions of superconducting material separated by thin insulating barriers,\cite{RMP} but in our case these islands will form spontaneously in a uniform material. In a loose sense, superconducting islands may be regarded as the opposite of vortices, but their length scale is much larger. In appropriate situations, these islands form a periodic pattern. We shall see that the rearrangement of these islands is related to the appearance of Hall voltage.

\section{SELECTION OF THE PROBLEM}
The situation we consider is as follows: the applied magnetic field is kept fixed at $0.5H_{c2}(0)=0.25\Phi_0/\pi\xi^2(0)$, where $\Phi_0$ is the quantum of flux and $\xi(T)$ is the coherence length at temperature $T$. At this field, the sample is in the mixed state. Initially, there is no net current and the situation is static. Then the current in the $x$-direction is gradually increased, until the film becomes normal. If the current increases sufficiently slowly, we may argue that we have a quasistationary situation and thus evaluate the properties of the superconductor as functions of the current.

The boundary conditions that are usually assumed in the TDGL treatment are 
continuity of the magnetic field and $\hat{\nu} \cdot \left( -i \nabla 
-{\bf A}\right) \psi =0$, where $\hat{\nu}$ is a vector perpendicular to the superconductor-insulator interface. However, this condition implies that the electric field is parallel to the interface, and I therefore suspected that it might not be appropriate for the study of the Hall voltage. Instead, I used the refined boundary condition suggested in Eq.~(8.26) of Ref.~\onlinecite{CS} at the boundaries $y=0$ and $y=d$. It turned out, however, that the same results are obtained without this refinement. At the electrodes ($x=0$ and $x=L$) I have considered two different boundary conditions. One case was that of periodic boundary conditions, which are frequently used to mimic an infinite sample in the $x$-direction; the other case was the Dirichlet condition, as appropriate for normal electrodes in which superconductivity is strongly suppressed. Periodic boundary conditions are not physically justified, but they lead to results that are simpler to analyze. Therefore, for exposition purposes, they will be presented first. A detailed discussion for the justification of the choice of the boundary conditions will be presented elsewhere.

In order to keep just a small number of parameters, Eq.~(\ref{ohm}) neglects the force exerted by the magnetic field on the normal electrons. Therefore, the Hall voltage that we obtain is that of the superconducting electrons only. Since the Hall field is much smaller than the field in the direction of the current, the Hall voltage should be, to a good approximation, the superposition of contributions from both kinds of electrons. We will also assume that the density and mobility of the normal electrons are not significantly affected by superconductivity. For a treatment that considers the entire resistivity tensor see, e.g., Ref.~\onlinecite{Fortin}.

We integrate the TDGL equations by means of a finite difference method, using essentially the same program as in Ref.~\onlinecite{Bolech}. In this method the sample is represented by a rectangular grid, consisting of $N_x \times N_y$ cells with spacings $a_x=L/N_x$ and $a_y=d/N_y$. Discrete values of $\psi$ are defined at every vertex and values of $A_{x,y}$ are defined at every link in the respective $x$ or $y$ direction. When periodic boundary conditions are used, they are imposed both on $\psi$ and on $\exp(iA_{y}a_{y})$. A standard initial state was obtained by raising the applied field from 0 to $0.5H_{c2}$ and then keeping the field fixed until a stable (or metastable) state was reached. The initial state for every run was then obtained by adding to $\psi$ at every vertex a complex random number with normal distribution, zero average and standard deviation 0.1. In a few cases, different histories were used.

The current enters the algorithm through the effect it produces: it raises the value of the magnetic field at one interface and lowers it at the other.

Most of the numerical studies in TDGL consider $\eta>1$, corresponding to cases in which TDGL can be derived from microscopic models; we have found that the effects reported here appear for $\eta$ significantly smaller than 1, as is the case for clean superconductors. 

\begin{figure}
\scalebox{0.85}{\includegraphics{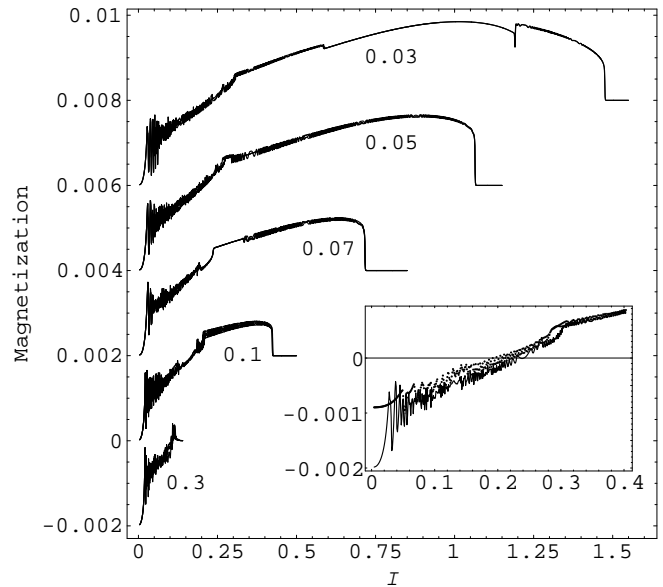}}%
\caption{\label{magnet}Clusters of curves of the magnetization of the film versus the current, with $\eta$ as a parameter. Each curve in a cluster has an initial state slightly different than those of the others. The magnetization is the volume average of the induced magnetic field, in units of $H_{c2}(0)$, divided by $4\pi$. The unit of current is $c\Phi_0/[2\pi\xi(0)]^2$ per cm of length in the $z$-direction. For each cluster there is a different value of $\eta$, which is marked next to it. Typically, each cluster contains four curves. For visibility, most clusters have been shifted in the vertical direction. Each curve starts at the current $4\times 10^{-3}$. At the right extreme of every cluster the film is in the normal state and the magnetization vanishes. The other parameters used in the calculations are: $N_x=80$, $N_y=16$, $a_x=a_y=0.5$, $\kappa=2$, $T=0.5$ and the size of the noise is the same as in Refs.~\onlinecite{Kato,Bolech}. The current increment between consecutive steps is $\Delta I=5\times 10^{-7}$ and the time increment varies from $\Delta t=0.008$ for $\eta=0.3$ to $\Delta t=0.0015$ for $\eta=0.03$. The applied field was kept fixed at $0.5H_{c2}(0)$. In the inset, one of the curves for $\eta=0.03$ is compared with the values obtained when the side of the cells in the calculation grid is decreased by a factor of 2 and the initial state is significantly different.}
\end{figure}

\section{RESULTS}
\subsection{Periodic boundary conditions}
We report on samples of thickness $d=8\xi(0)\approx 7\xi(T)$. For these samples the standard initial state had a row of vortices at $y=d/2$ and the average distance between consecutive vortices was $2.5\xi(0)\approx 2.2\xi(T)$.

Figure \ref{magnet} shows the magnetization $M$ of the film as a function of the current $I$, for several values of $\eta$. The curves start at $I=4\times 10^{-3}$ rather than $I=0$ in order to chop off the influence of the random numbers added to the initial state. The general behavior is similar for all the curves (including additional values of $\eta$ not shown in the figure). For every curve there is a small region $0\le I\le I_1$ where $M$ is a smooth function of $I$. For $I>I_1$, there are points where the slope changes discontinously; at some of these points the slope changes sign and sometimes there is just a kink. (The curves themselves have to be continuous if $I$ changes at a finite rate.) The points where the slope is discontinuous coincide with the entrance or exit of vortices in and out of the film. The precise points where these discontinuities occur vary among different runs, but the general behavior is always the same. There is a current $I_2>I_1$ where the magnetization changes from diamagnetic to paramagnetic. If we disregard the rapid oscillations of the curves and consider only their smoothed trends, we observe that there is a current $I_3>I_2$ where the slope decreases pronouncedly. (For $\eta=0.03$, $I_3=0.31$.) Finally, there is a current $I_\c$ where the film becomes normal and the magnetization drops to zero. We also observe that there are regions, like a region that contains $I=I_3$ for $\eta=0.07$, where oscillations are practically absent and all the curves in the cluster coalesce. One might be tempted to suspect that in this regime vortices do not enter or leave the film, but closer examination shows that large regions of the sample have become normal (typically, $|\psi|<10^{-3}$ in these regions); under this condition, the concept of vorticity loses its significance.

\begin{figure}
\scalebox{0.85}{\includegraphics{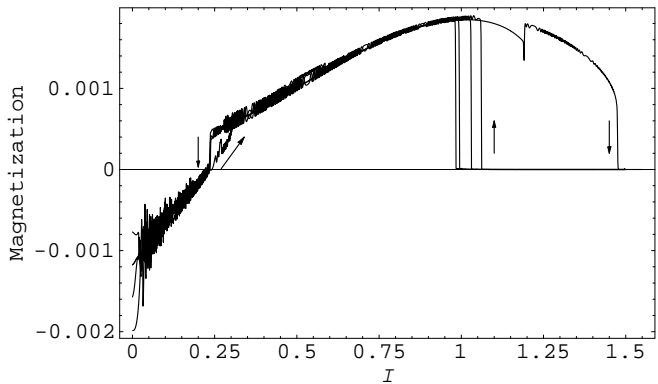}}%
\caption{\label{reverse}$M(I)$ for $\eta=0.03$. One curve is for 
increasing current, as in Fig.~\ref{magnet}, and four curves are for decreasing current. The other parameters are as in Fig.~\ref{magnet}. }
\end{figure}

$I_\c$ is the limit of metastability of the superconducting state, but at these high currents also the normal state is metastable. Figure~\ref{reverse} shows a cluster of $M(I)$ curves for a case in which the current was decreased from above $I_\c$ to 0.  We see that there is a limit of metastability $I_{c\downarrow}\sim 0.98$ and the sample switches to the superconducting state when this limit is approached. The curves $M(I)$ are roughly the same when $I$ is raised or lowered, except for a large hysteresis loop between $I_\c$ and $I_{c\downarrow}$, and a small hysteresis loop between $I_3$ and $I_{3\downarrow}\sim 0.24$. All the curves in the cluster undergo a drop in the magnetization at the same current $I_{3\downarrow}$. Hysteresis is a common phenomenon in superconductivity when the current is varied; a particularly well known case is that of an underdamped Josephson junction. 

In order to test the reproducibility of our results, we performed several runs with different computational parameters. The solid line in the inset of Fig.~\ref{magnet} is one of the lines in the cluster for $\eta=0.03$, whereas the dots correspond to essentially the same physical parameters, but the computing grid was denser ($N_x=159$, $N_y=32$) and the rate of change of the current was increased by a factor of 5 ($\Delta t=0.0003$). In addition, the initial state was significantly different from the standard one; the average distance between consecutive vortices was $2.2\xi(0)$ rather than $2.5\xi(0)$. Due to the different initial state, the initial magnetization for this exceptional run is about half that of the runs in the cluster; however, for $I>I_3$, it appears that the memory about the initial state has been lost.

\begin{figure}
\scalebox{0.85}{\includegraphics{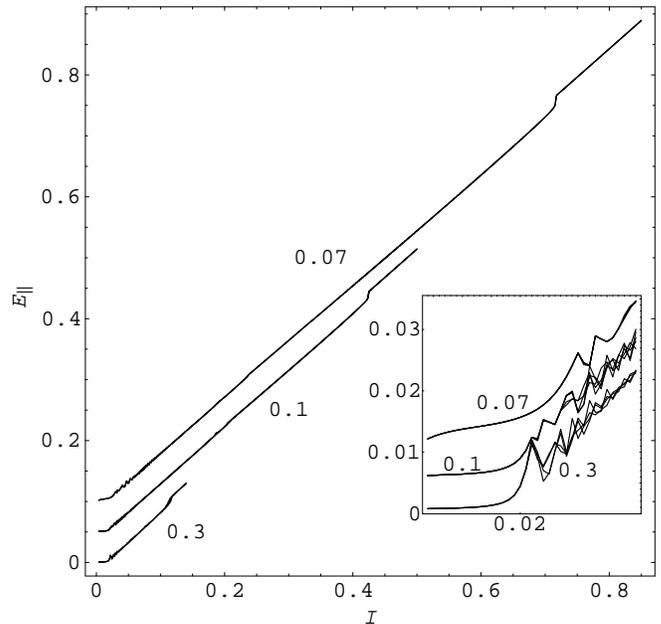}}%
\caption{\label{volt}Clusters of curves of the parallel component of the electric field versus the current, with $\eta$ as a parameter. The electric field unit is $c\Phi_0/[8\pi^2\kappa^2\sigma\xi^3(0)]$, where $\sigma$ is the (normal) conductivity of the sample. For visibility, the upper clusters have been shifted in the vertical direction. In this graph each line starts at $I=E_\parallel=0$. The other parameters used are the same as in Fig.~\ref{magnet}. The inset shows the initial part of the considered curves.}
\end{figure}

\begin{figure}
\scalebox{1.2}{\includegraphics{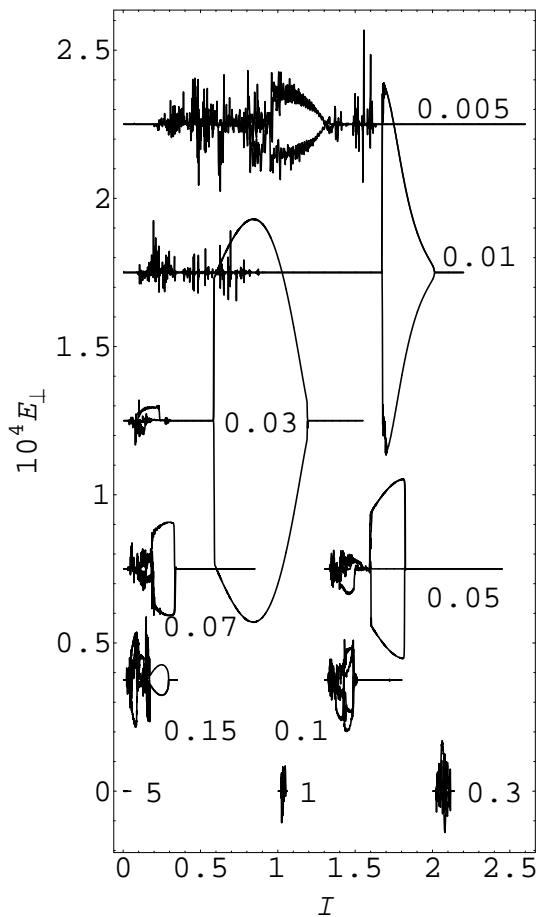}}%
\caption{\label{Hall}Clusters of curves of the Hall component of the electric field versus the current, with $\eta$ as a parameter. Most clusters have been shifted in the vertical and/or horizontal direction, but they are all in the same scale. Each cluster starts at $I=4\times 10^{-3}$ and at the right extreme $E_\perp=0$. The other parameters used are the same as in Fig.~\ref{magnet}, except for $\Delta I$, which was taken as $10^{-6}$ for $\eta=0.01$ and $\eta=0.005$, and as $2.5\times 10^{-7}$ for $\eta=0.15$. $E_\perp$ usually vanishes, but for some regions of $I$ $E_\perp\neq 0$. In these regions $E_\perp$ is chaotic: for some runs is positive and for others is negative. For $0.01\le\eta\le 0.07$ there is a region where $E_\perp(I)>0$ for some lines and $E_\perp(I)<0$ for others, but $|E_\perp(I)|$ is the same for all the lines in the cluster, so that a ``bubble" is formed. Note that the bubble for $\eta=0.03$ invades the neighboring clusters.}
\end{figure}

Figure \ref{volt} shows the average electric field $E_\parallel$ in the direction of the current, for a few values of $\eta$. These results were obtained by averaging the parallel component of the electric field over all the lines in the grid in the direction of the current, and also over 2000 consecutive time steps. The results resemble those obtained for long channels in the absence of applied field.\cite{resistive} For $I<I_1$, $E_\parallel$ is very small and for $I>I_\c$ (normal state) we obtain $E_\parallel=2\kappa^2 I/d$, which is Ohm's law in the units we are using. In the intermediate region, except for small oscillations, we obtain straight lines with the same slope as in the normal state.

The inset in Fig.~\ref{volt} is a close up for low currents. The reason that $E_\parallel$ doesn't vanish completely for $I<I_1$ is that our method of evaluation is not exactly stationary. For example, if we start from the point at $I=0.01$ for $\eta=0.07$ and keep the current fixed, $E_\parallel$ decays with a time constant of 18 time units. This result can be understood in terms of vortex motion: the current exerts a force on the vortices, which attempts to drive vortices into or out of the film. However, for $I<I_1$, the Bean-Livingston barrier\cite{barrier} prevents vortices from crossing the interface. Therefore, for constant currents the vortices attain equilibrium positions and stop moving; it is only the change in current that keeps the vortices in motion. Likewise, $E_\parallel$ increases close to $I_1$; this happens because the configuration becomes unstable and vortices accelerate.

Figure \ref{Hall} presents the average of the $y$-component of the electric field, $E_\perp$, for several values of $\eta$. This result was obtained by averaging this component over all the lines in the grid in the direction perpendicular to the current, and also over 2000 consecutive time steps. There are essentially two regions of currents for which $E_\perp$ does not vanish. There is a clearly distinguished feature at the right, present for $0.01\le\eta\le 0.07$, which we call a ``bubble"; there is also a less clear feature at the left, which we call a ``jitter." In these regions $E_\perp$ is chaotic: minute differences in the state of the system when it enters the region cause it to assume any of several very different functions $E_\perp(I)$. A major difference between these two regions is that for the bubble the system chooses among a small number of possibilities, which appear to be symmetric with respect to the line $E_\perp(I)=0$, whereas for the jitter the number of possibilities is large and has no obvious pattern. In most cases $E_\perp(I)$ is either positive or negative in the bubble region, but there are a few cases (as a case shown for $\eta=0.05$) in which $E_\perp(I)=0$. 

For $\eta=0.01$ the bubble ends at $I_\c$, but for $\eta\ge 0.03$ the bubble ends at a current smaller than $I_\c$. In the case $\eta=0.03$, the bubble region coincides with the central part of the region $I_3\le I\le I_\c$ in which the magnetization curve looks lower and smoother than its continuation at both sides. As $\eta$ increases, the bubble moves to the left, until it merges with the jitter.

In order to elucidate which part of the film gives rise to $E_\perp$, we have also evaluated its averages restricted to the sides of the cells that touch one of the interfaces. We found that $E_\perp$ in the bubble region is not influenced by the interface where the applied magnetic field is augmented by the induced magnetic field; $E_\perp$ in the jitter region is influenced by both interfaces.  

\begin{figure}
\scalebox{0.85}{\includegraphics{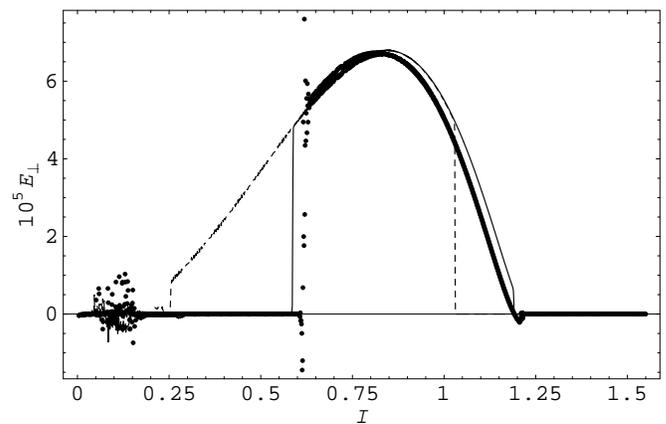}}%
\caption{\label{compare}Hall component of the electric field for different histories. The solid line and the dots describe the same runs as in the inset of Fig.~\ref{magnet}. In the absence of scattering, the dots look like a thick line. The dashed line describes a case in which the current was decreased.}
\end{figure}

\begin{figure}
\scalebox{1.0}{\includegraphics{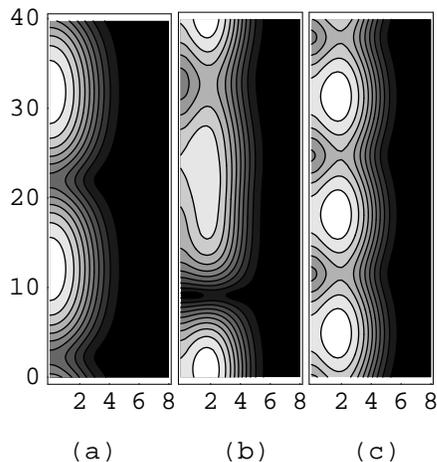}}%
\caption{\label{snap}Contour plot of the size of the order parameter $\psi$. $|\psi|$ is larger in the lighter areas. The current flows in the vertical direction. The unit of length is $\xi(0)$. Note that the scale is different for each direction. All the parameters are those of the dotted line in the inset of Fig.~\ref{magnet}. (a) $I=0.555$, slightly before the bubble. (b) $I=1.195$, almost at the end of the bubble. (c) $I=1.235$, slightly above the bubble region.}
\end{figure}

Since the Hall voltage we find might just be due to some instability of 
our numeric algorithm, we compare in Fig.~\ref{compare} the values of 
$E_\perp$ obtained in different ways. Two curves are for currents 
increasing with time, but for grids and time steps of different sizes, and 
very different initial states (the same runs as in the inset of 
Fig.~\ref{magnet}). Except for the amount of scattering and overshoot, the results coincide in the bubble region (although they have opposite signs in the jitter region). The third curve is for decreasing current. In spite of the large hysteresis found in Fig.~\ref{reverse}, this line also coincides with those of increasing current in the region where the Hall voltage is present. It should be mentioned that for decreasing current $E_\perp$ showed up in the bubble region only for about half of the runs. When it did show up, it appeared immediately with the switch to the superconducting state and remained present down to $I=I_{3\downarrow}$. No visible change in $E_\perp$ is obtained if our boundary condition at $y=0$ and $y=d$ is replaced by $\hat{y} \cdot \left( -i \nabla -{\bf A}\right) \psi =0$. 

We would like to gain some intuition concerning the reason for the existence of this chaotic Hall voltage. For this purpose, we have mapped the size of the order parameter for several currents. Figure~\ref{snap}(a) is a contour plot of $|\psi|$ for a current slightly below the bubble region. This current is already large enough to turn into normal the entire area close to the interface where the magnetic field is large. The most interesting feature is that the superconducting region is not just a stripe parallel to the current direction, but it rather concentrates into a discrete set of superconducting islands; for the current in Figure~\ref{snap}(a), there are two such islands. Instead, for $I=1.235$, slightly above the bubble region, we find that there are three well defined islands. In the bubble region itself, a process occurs in which two islands have to turn into three; during this process the islands assume an irregular shape, as in Fig.~\ref{snap}(b). This deformation imposes an overall motion of the superconducting part either in the direction of the current or against it, and this motion produces the Hall voltage.

The size of the Hall voltage in the bubble region does not depend on the rate at which the current is swept. Increasing this rate by a factor of 40 just produces some extra overshooting, but near its maximum $|E_\perp (I)|$ looks essentially the same as in Fig.~\ref{compare}. Moreover, if we stop increasing $I$ and keep it at a fixed value in the bubble region, $E_\perp (I)$ remains constant in time, except for small oscillations that might be due to numeric inaccuracy. According to our interpretation, this means that when $I$ is in the bubble region the islands have to move, either in or against the direction of the current, even if $I$ is kept unchanged.

\begin{figure}
\scalebox{0.85}{\includegraphics{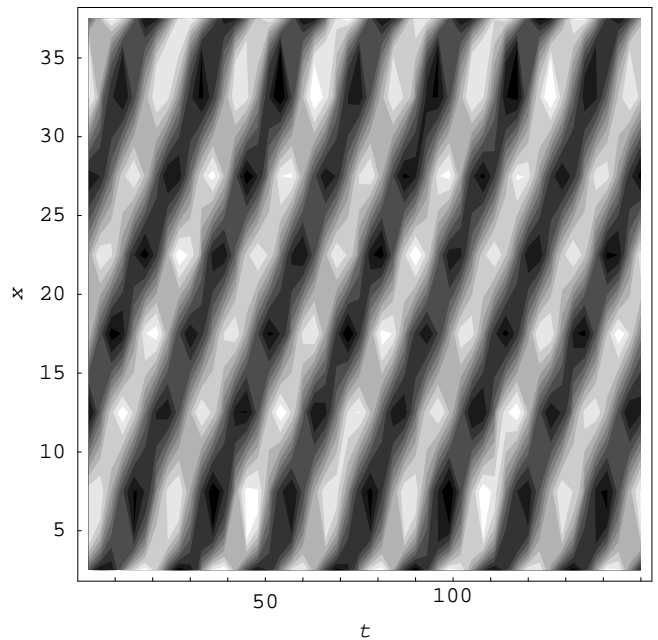}}%
\caption{\label{wave}Contour plot of $E_\perp (x,t)$. The unit of time 
is $4\pi\kappa^2\sigma\xi(0)^2/c^2$. In order to evaluate 
$E_\perp (x,t)$, the sample was divided into 8 fringes, perpendicular to 
the current, and $E_\perp$ was averaged on each fringe. For the case described here, $\eta=0.03$, $I=0.85$, and the other parameters are as in Fig.~\ref{magnet}.  The regions $x<2.5$ and $x>37.5$ have been chopped off, due to the finite width of the fringes. This contour plot indicates that  $E_\perp (x,t)$ behaves as a one-dimensional wave, with wave velocity $0.7c^2/4\pi\kappa^2\sigma\xi(0)$. }
\end{figure}

It is easy to monitor this motion by following the time dependence of $E_\perp$. The effect of the motion is not observed if we take the average of $E_\perp$ along the entire range $0\le x\le L$, but becomes visible if we divide the sample into fringes, perpendicular to the current direction. Figure~\ref{wave} is a contour plot for $E_\perp (x,t)$, which shows the motion of the maxima and minima of $E_\perp$. Since these maxima and minima are expected to move together with the islands, their velocity should be equal to that of the islands. For the case described in Fig.~\ref{wave}, the islands move with a velocity of $0.7c^2/4\pi\kappa^2\sigma\xi(0)$.

\begin{figure}
\begin{tabular}{r}
\scalebox{0.83}{\includegraphics{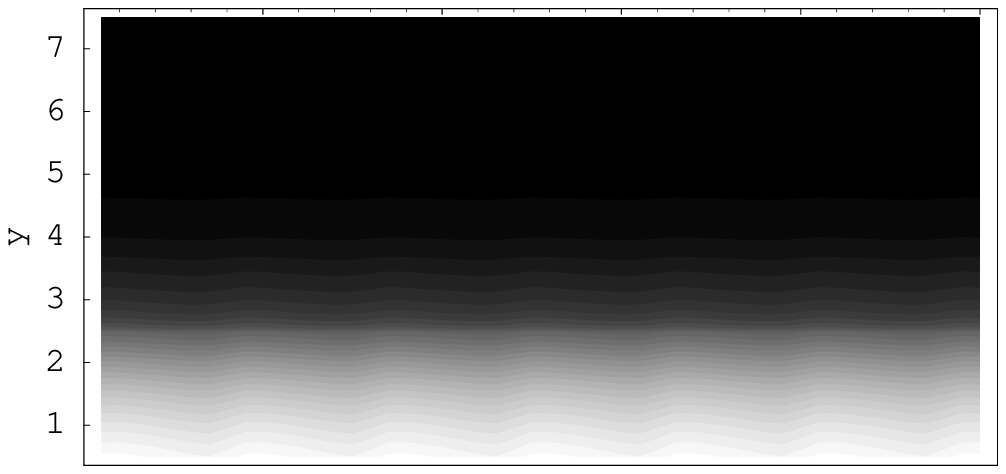}} \\
\scalebox{0.85}{\includegraphics{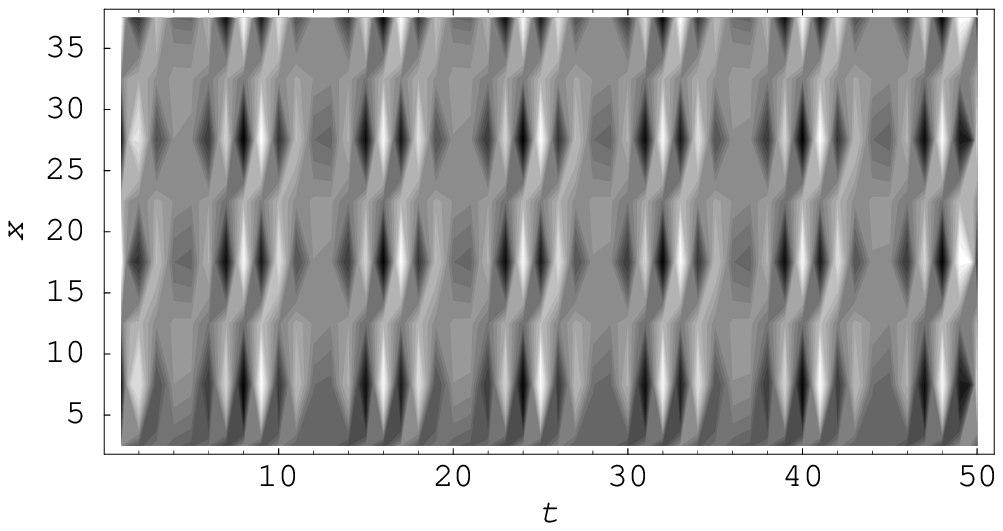}} 
\end{tabular}
\caption{\label{xy}Distributions of the superconducting electrons in the $x$- and $y$-directions, as functions of time. $\eta=0.03$, $I=0.4$, and the other parameters are as in Fig.~\ref{magnet}. The size of the binnings for $x$, $y$ and $t$ are 5, 1 and 1, respectively.}
\end{figure}

A quantity that is easier to interpret is the ``number of superconducting 
electrons." By defining $n_s^x(x,t)=\int |\psi(x,y,t)|^2\,dy$ and 
$n_s^y(y,t)=\int |\psi(x,y,t)|^2\,dx$ and drawing contour plots of these 
quantities, we can visualize how the superconducting regions move. We have 
applied this procedure for the case $\eta=0.03$ while the current is kept constant at $I=0.4$; for this current $E_\perp$ vanishes on the average. The upper panel in Fig.~\ref{xy} shows that the superconducting regions do not move in the $y$-direction. The lower panel should be compared with Fig.~\ref{snap}(a): while in Fig.~\ref{snap}(a) there are only two islands, the distribution of $n_s^x$ exhibits four fringes. (Half a fringe was chopped off by the binning.) Examination of this lower panel shows that the islands switch positions with a period of about two time units, i.e., the centers of the islands become the valleys that separate between them, and vice versa. There is an additional phenomenon, with a period of about 8 units, in which the islands are blurred. In view of Fig.~\ref{xy}, Fig.~\ref{snap}(a) should be understood as an average over a period of time that is neither too long nor a multiple of these periodic processes.

From Fig.~\ref{xy} we also learn that $E_{||}$ should not be interpreted as due to motion of the islands in the $y$-direction. Rather, it acts like a Josephson field that causes oscillatory motion of the superconducting regions in its own direction. When the volume average of $E_{||}$ is analyzed as a function of time, we find that is almost constant, with a small alternating part with period 8.

\subsection{Dirichlet condition}

\begin{figure}
\scalebox{0.85}{\includegraphics{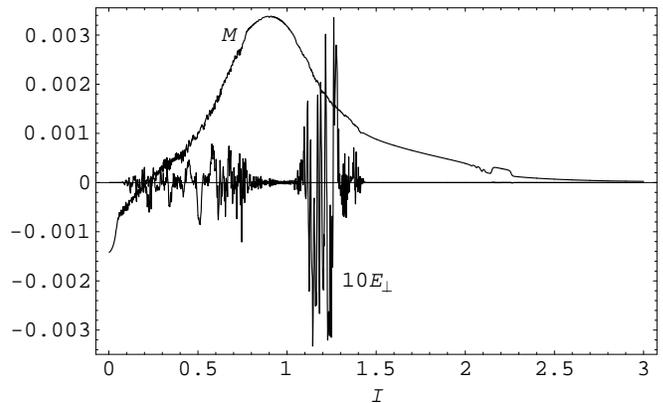}}%
\caption{\label{DHall}Magnetization (curve that looks like a mountain in the horizon) and Hall component of the electric field (curve that oscillates strongly about zero), for Dirichlet boundary conditions. Only one run was carried out. The size of the sample is $L=80\xi(0)$, $d=16\xi(0)$. Other parameters: $\eta=0.03$, $\kappa=2$, $T=0.5$, $a_x=a_y=0.5\xi(0)$, $\Delta I=10^{-6}$, $\Delta t=0.0015$.}
\end{figure}

\begin{figure}
\begin{tabular}{ccc}
\scalebox{0.6}{\includegraphics{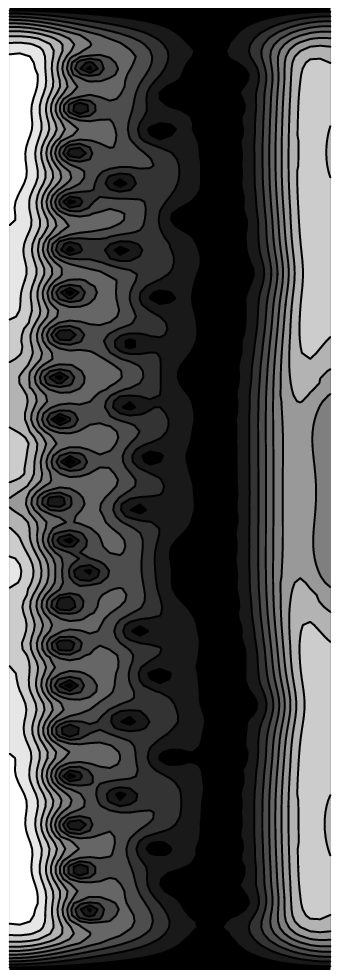}} &
\scalebox{0.6}{\includegraphics{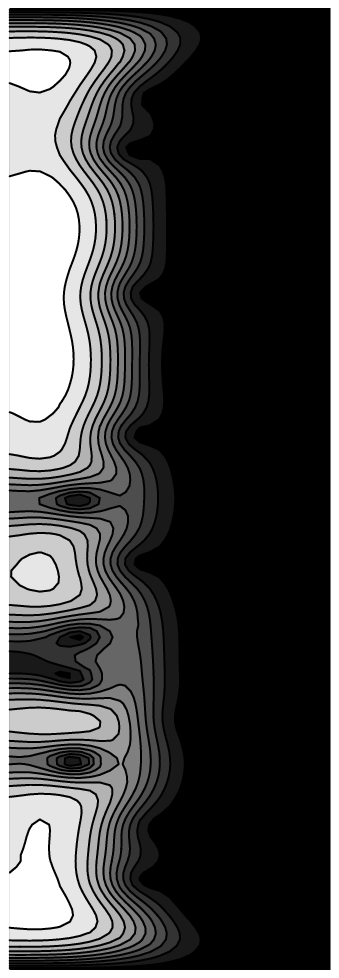}} &
\scalebox{0.6}{\includegraphics{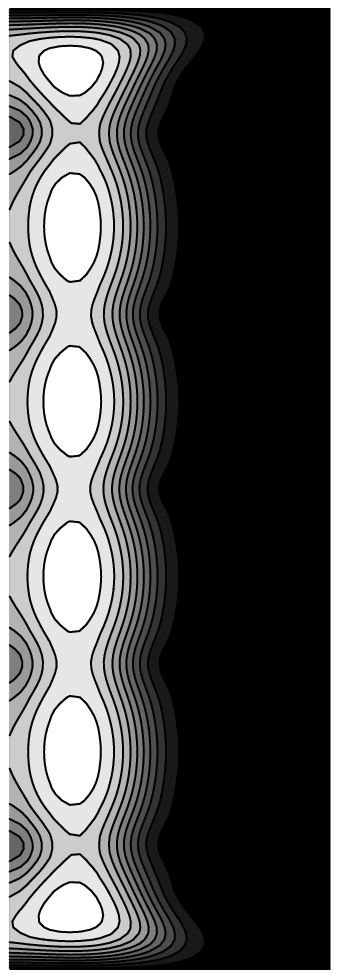}} \\
(a) & (b) & (c) \\
\scalebox{0.6}{\includegraphics{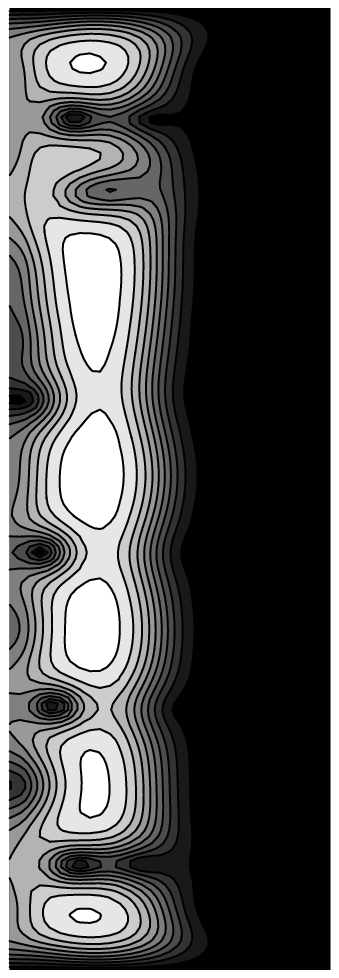}} &
\scalebox{0.6}{\includegraphics{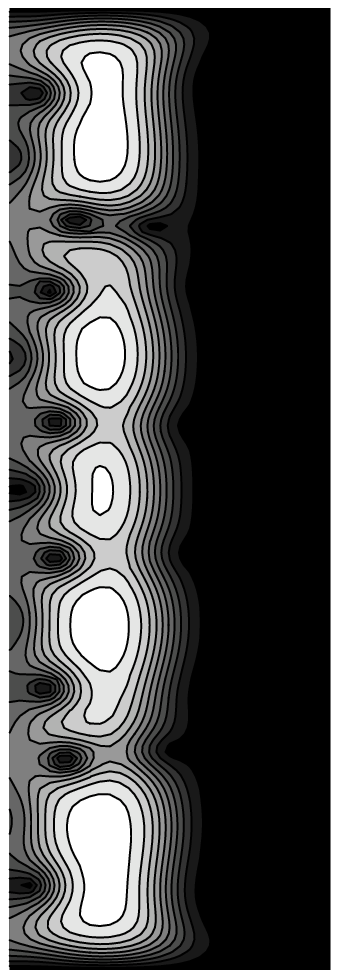}} &
\scalebox{0.6}{\includegraphics{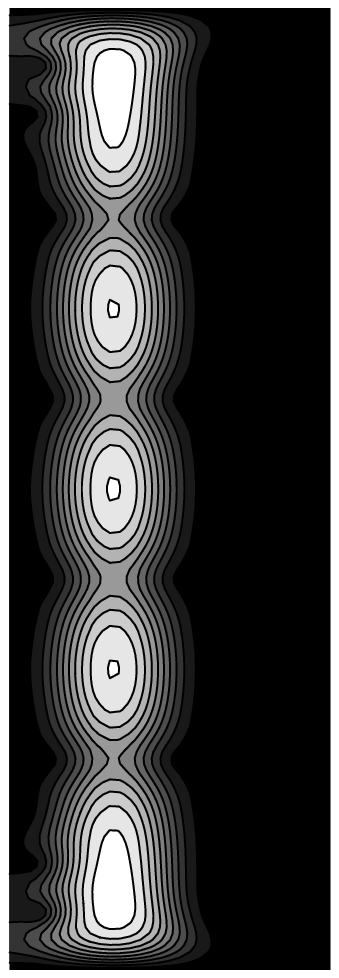}} \\
(d) & (e) & (f)
\end{tabular}
\caption{\label{fixed}Contour plot of the size of the order parameter $\psi$ for the process described in Fig.~\ref{DHall}. (a) $I=0.08$. (b) $I=0.75$. (c) $I=1$. (d) $I=1.2$. (e) $I=1.3$. (f) $I=1.5$. }
\end{figure}

In the previous case the islands formed a perfectly periodic structure, due to the artificial requirement of periodic boundary conditions in the direction of the current. We may still anticipate that for any reasonable boundary condition at the electrodes there will be a finite number of islands and the creation of a new island will involve distortion. As a more realistic boundary condition at the electrodes, we considered Dirichlet conditions. The boundary condition for {\bf A} was the assumption that the electric field in the electrodes is always in the $x$-direction, implying that $A_y$ remains fixed in time. 

The results are complicated by the fact that now the electrodes pin the superconducting islands and impede their free motion. For $L=40\xi(0)$ the influence of pinning is so strong that we did not find island fragmentation. We may expect that the influence of pinning will be weaker for longer samples, but, if the size of computational cells is kept smaller than the coherence length, larger $L$ implies heavier numeric calculations.
As a compromise, we studied cases where $L=80\xi(0)$; we also doubled the thickness of the sample, although that was not really necessary. For these thicker samples the initial state had three rows of vortices.

Figure~\ref{DHall} shows the magnetization and the average Hall field as functions of the current.
We still obtain that $E_\perp$ vanishes for small and large currents and is present in an intermediate region. In this region there are two subregions where $E_\perp$ is significant, separated by a ``quiet" subregion close to $I\sim 1$. For $1.05\alt I\alt 1.45$, $E_\perp$ has the same order of magnitude as in the ``bubble" for periodic boundary conditions, but is now of the ``jitter" type.

Figure~\ref{fixed} shows the shape of the order parameter as the current is increased. For $I=0.08$ there are three rows of vortices in the sample, but around the row at the right $\psi$ is too small to make them visible. For this row the value of $|\psi|$ at the saddle points is of the order of 0.02. In a rough sense, there is a symmetry mirror at $x=L/2$ and, accordingly, we see that $E_\perp$ vanishes. For $I=0.75$ it would be useless to look at vortices; the entire area at the right has become normal and the notion of superconducting islands becomes more meaningful. There is no symmetry with respect to $x=L/2$ and, accordingly, $E_\perp$ is large. For $I=1$ a clear pattern of islands has developed. Two border islands are pinned at the electrodes and four islands in the middle form a nearly periodic pattern. This pattern is symmetric with respect to $x=L/2$ and $E_\perp$ is small. For $I=1.5$ we have again a regular symmetric pattern of islands and $E_\perp=0$. In contrast with the previous subsection, ``counting" islands is not clear, since the border islands may be regarded as fractions that do not remain fixed. The cases $I=1.2$ and $I=1.3$ illustrate the passage between the regular situations at $I=1$ and $I=1.5$. During this passage the regions where $\psi$ is significant move back and forth and assume irregular shapes. Accordingly, $E_\perp$ is large and changes sign. For large currents the total magnetic field is large at both superconducting-insulating boundaries (with opposite signs). Figure~\ref{fixed}(f) shows that the superconducting islands take refuge along a stripe where the magnetic field is small, and this enables them to survive up to large currents.

\begin{figure}
\scalebox{0.85}{\includegraphics{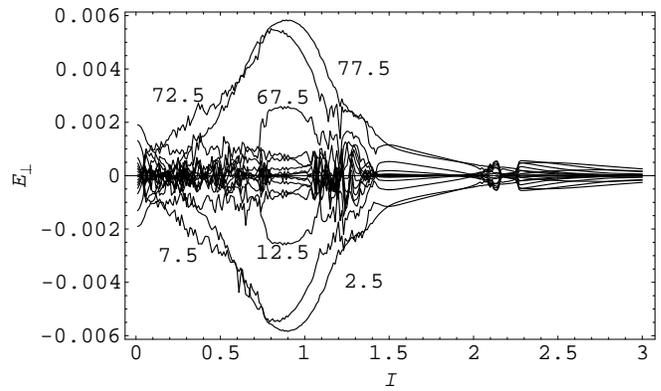}}%
\caption{\label{all}The sample described in Fig.~\ref{DHall} was divided into 16 fringes, perpendicular to the current. The curves in the graph describe the average of $E_\perp$ in each of these fringes. The number next to some of the curves is the value of $x$ at the middle of the fringe. The curves for fringes close to $x=40\xi(0)$ are hard to follow and have not been marked.}
\end{figure}

Contrary to the previous subsection, the islands are now blocked by the electrodes and cannot move freely in or against the direction of the current; therefore, $E_\perp$ roughly vanishes on the average. However, islands might conceivably be formed near the middle of the sample and could migrate to the electrodes and disappear at them. If this is the case, we would expect $E_\perp(x)$ to differ from zero when averaged over fringes of limited width. Figure~\ref{all} shows the averages of $E_\perp$ over fringes of width $\Delta x=5\xi(0)$. We see that $E_\perp(x)$ is roughly antisymmetric with respect to  the line $x=40\xi (0)$ at the middle of the sample. We also see that $E_\perp (x)$ does not vanish in regions where the average $E_\perp$ does ($I<0.1$ and $I>1.5$). The difference between these and the other regions is determined by whether the antisymmetry with respect to the middle of the sample is perfect or not. In the following subsection we will check whether the conjecture of island migration describes an appropriate scenario.

In the regions where the average $E_\perp$ does not vanish and the islands are irregular, $E_\perp (x)$ is chaotic. Since this effect is very similar to the case that will be discussed in the following subsection, we do not provide a figure for it.

\subsection{Nonuniform sample\label{GG}}
It turns out that the results are more clear cut if the electrodes are ``smeared" by gradually reducing superconductivity in their proximity.

A possibility for describing locally stronger or weaker superconducting materials is the replacement of Eq.~(\ref{psi}) by
\be
{\partial_t \psi }=-\frac 1\eta \left[ \left( -i\nabla -%
{\bf A}\right) ^2\psi +\left( 1-T\right) \left( |\psi |^2-1-\delta\right) \psi
\right] +\tilde{f} \;, \label{delta} 
\ee
where $\delta$ is a function of position. If $\delta>0$ (respectively $\delta<0$) in some place, superconductivity is stronger (respectively weaker) at that place; the case $\delta=-1$ describes the situation in which the critical temperature has been reached.

We considered a sample of thickness $d=8\xi(0)$ and length $L=120\xi(0)$. Its central segment $20\le x\le 100$ was uniform ($\delta =0$), but close to the electrodes ($0<x<20$ and $100<x<120$) $\delta$ varied linearly with $x$, reaching $\delta =-1$ at the electrodes. In this way, the normal material at the electrodes was met when superconductivity had already disappeared.

\begin{figure}
\scalebox{0.85}{\includegraphics{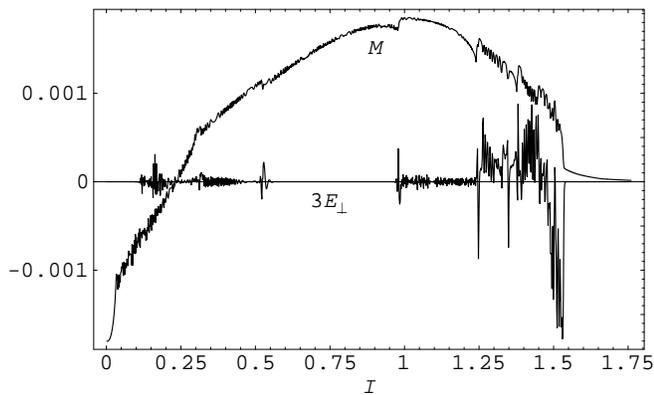}}%
\caption{\label{gradual}Magnetization and average Hall component of the electric field for the sample considered in subsection \ref{GG}. The averages of $M$ and $E_\perp$ were taken over the uniform central part ($20\le x\le 100$) only. Other parameters: $d=8\xi(0)$, $\eta=0.03$, $\kappa=2$, $T=0.5$, $a_x=a_y=0.5\xi(0)$, $\Delta I=8\times 10^{-7}$, $\Delta t=0.0015$.}
\end{figure}

Figure~\ref{gradual} shows the results for the magnetization and for the average Hall field. Comparison of this figure with Fig.~\ref{DHall} shows that we have recovered some of the features encountered for the case of periodic boundary conditions: there is an intermediate region where the Hall field vanishes and there is only a limited region where this field is large. Also, there is a current where the over-all slope of the magnetization curve decreases significantly. On the other hand, there is no extended region where the Hall field preserves its sign.

\begin{figure}
\scalebox{0.85}{\includegraphics{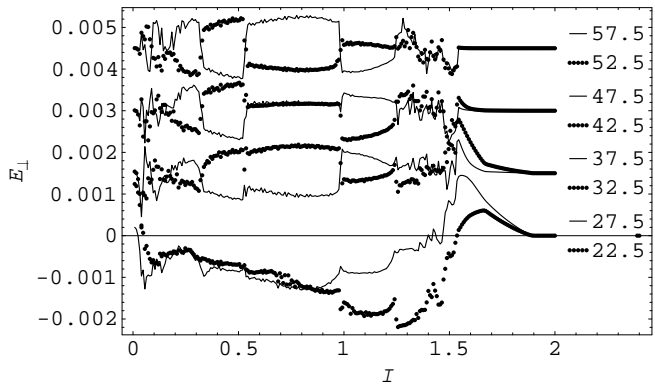}}%
\caption{\label{pairs}Hall field for different positions in the uniform segment of a nonuniform sample. $E_\perp$ has been averaged over fringes of width $5\xi(0)$. Each curve is marked by the distance $x$ between the middle of the fringe and one of the electrodes.  For $x>60$, $E_\perp$ can be obtained from $E_\perp (x)\approx -E_\perp(120-x)$. For visibility, the results were grouped into pairs and the upper pairs were shifted in the vertical direction. At $I=2$,  $E_\perp$ vanishes for all fringes. All parameters are the same as in Fig.~\ref{gradual}. }
\end{figure}

As in the previous subsection, the average Hall field is just a residual effect, due to its antisymmetry as a function of position. In Fig.~\ref{pairs} we present $E_\perp$ for several values of $x$ within the segment where the sample is uniform. Since antisymmetry is quite well obeyed, these results are shown for half of the sample only.

\begin{figure}
\scalebox{0.85}{\includegraphics{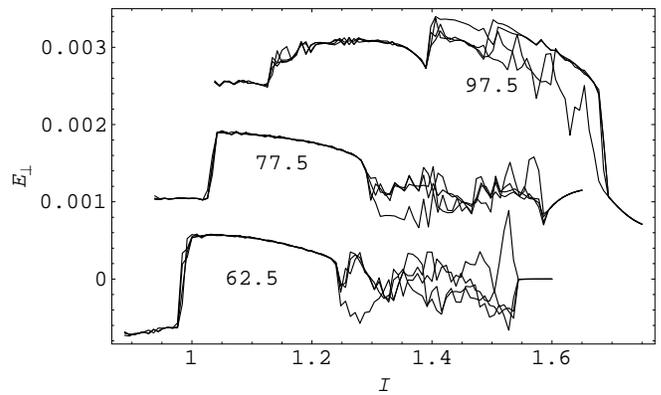}}%
\caption{\label{allC}Hall field averaged over fringes of width $5\xi(0)$ for miscellaneous values of $x$ and for a range of currents that contains the region where the average over $x$ is large. All parameters are the same as in Fig.~\ref{gradual}, but there are clusters of curves starting at four different initial states. For visibility, the cluster for $x=77.5$ has been shifted by 0.05 to the right and 0.0012 upwards; the cluster for $x=97.5$ has been shifted by 0.15 to the right and also 0.0012 upwards.}
\end{figure}

In order to check whether $E_\perp (x)$ is chaotic, we repeated the calculation of Fig.~\ref{pairs} for several initial states. Some of the clusters of curves obtained are shown in Fig.~\ref{allC}. We see that in the regions where the average value of $E_\perp$ is small in Fig.~\ref{gradual} all curves practically coalesce, whereas in the region $1.25\alt x\alt 1.5$ there is chaotic behavior.

\begin{figure}
\scalebox{0.85}{\includegraphics{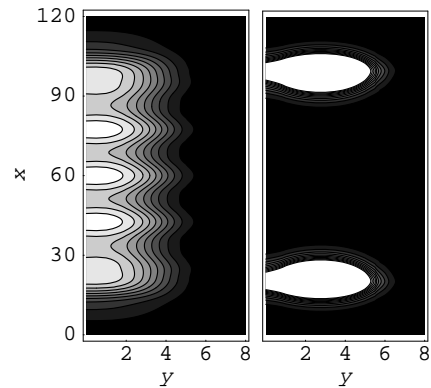}}%
\caption{\label{psG}Contour plots for the size of the order parameter for $I=0.88$ (left) and $I=1.76$ (right). The other parameters are as in Fig.~\ref{gradual}. }
\end{figure}

As in the previous cases, we would like to relate the behavior of the Hall field to the shape of $|\psi|$. Figure~\ref{psG} shows these shapes for currents where the average Hall field vanishes, before and after the region where this average is large. As might have been expected, these shapes are symmetric with respect to the transformation $x\rightarrow 120-x$. A striking feature is that for $I=1.76$ superconductivity concentrates into two sharply bounded islands at the borders between the segment where the material is uniform and those where superconductivity gradually decreases. These islands may be regarded as a generalization of surface superconductivity and are probably the reason for the long tail of the magnetization in Fig.~\ref{gradual}.

\begin{figure}
\scalebox{0.85}{\includegraphics{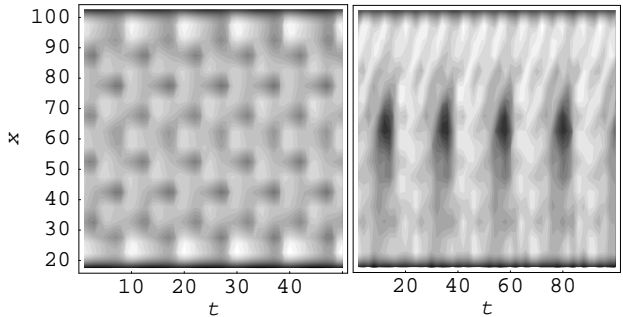}}%
\caption{\label{breathing}Contour plots for the average density of superconducting electrons along fringes perpendicular to the current, for the sample considered in subsection \ref{GG}. For distances smaller than $15\xi(0)$ from an electrode, $|\psi(x,t)|^2$ is very small and these regions are not shown. Left: $I=0.88$. The size of the binning along the $t$-axis is 1, and along the $x$-axis is 5. Right: $I=1.32$. The size of the binning along the $t$-axis is 2, and along the $x$-axis is 5. }
\end{figure}

We now attempt to find a relation between $E_\perp$ and island motion. As in the lower panel of Fig.~\ref{xy}, the contour plots in Fig.~\ref{breathing} describe the density of superconducting pairs, $|\psi(x,t)|^2$, averaged over fringes of width $\Delta x=5\xi(0)$. The plot at the left is for a current such that $E_\perp =0$ in Fig.~\ref{gradual} and at the right for a current such that $E_\perp$ is large. The plot at the left looks like a standing wave superimposed on a nonuniform background. This means that there is no net motion of the islands, and the fact that $E_\perp (x,t)$ does not vanish for every value of $x$ should be attributed to a nonlinear effect. The plot at the right might be described as a case of ``breathing." The most notable feature of this plot is that $|\psi(x,t)|$ decreases and increases periodically for $x\sim 65$. In addition, we see that the island-pattern moves asymmetrically: there is no net motion for $x<50$, whereas there is motion away from the center for $x>70$. This confirms the conjecture that the existence of an average Hall field is related to island motion.

\section{DISCUSSION}
We have found that for appropriate parameters a nearly periodic pattern of superconducting islands can form in a superconducting sample in the Hall configuration. There are situations for which a pattern becomes unstable and the islands become irregular, until a new regular pattern is attained. While the islands are arranged irregularly, they move in (or against) the direction of the current and the contribution of the superconducting electrons to the average Hall voltage does not vanish. Even when the islands do not move in a preferred direction, the size of the order parameter oscillates in time, so that local maxima of the order parameter may periodically become local minima.

This scenario shows up very clearly when periodic boundary conditions are imposed at the electrodes. For other boundary conditions, these phenomena are still qualitatively present, but become more difficult to analyze. The difference between realistic and periodic boundary conditions is that the latter don't pin the superconducing islands.

We might mimic a situation in which islands are not pinned by taking a 
sample which is infinitely long in the direction of the current flow; 
however, we may expect that no Hall voltage would appear in this limit, 
since the islands structure could shrink gradually and no distortion would be required. It seems therefore that some amount of pinning is desirable. This is consistent with the theoretical results obtained in Ref.~\onlinecite{Fortin}: the strongest oscillations of the Hall voltage were obtained for an intermediate amount of defects. In the calculations of Ref.~\onlinecite{Fortin} the defects were assumed to be located in a periodic array; this arrangement should be expected to favor the formation of periodic arrays of islands. Artificially created periodic arrays of defects in superconductors have been available for a long time.\cite{sch} We also found that gradual decrease of the superconducting strength of the superconducting material near the electrodes may help to create conditions similar to the periodic case.

One might argue that there is no qualitative difference between the formation of a new island and the entrance of a vortex: when a vortex enters the sample, the others have to move away from it and this motion gives rise to Hall voltage. There is, however, a quantitative difference: vortices feel each other over distances of the order of the coherence length, whereas islands seem to feel each other over distances that are larger by at least an order of magnitude.

We do not have a tested hypothesis concerning the physical parameters that control the size of the islands. Conceivably, the length of the islands could be proportional to their width and this could be controlled by the width of the region where the magnetic field is sufficiently small. On the other hand, since the effects described in this article have only been found for small values of $\eta$, we might suspect that the size of an island is related to the penetration length of the electric field, $\xi(T)/\sqrt{\eta}$.\cite{Kop}

Our results are indirectly supported by the experiments in Ref.~\onlinecite{Fortin}. In that reference, which used a particularly clean material (so that $\eta$ was indeed expected to be small), the variable quantity was the applied field, whereas the current was fixed. However, the Hall voltage exhibits the same qualitative features as in our case: it is present in limited regions only, is absent close to the critical current, exhibits oscillations and its sign is reversed for different experiments that were apparently identical. 

The fact that we do obtain a Hall voltage using Eq.~(\ref{psi}) challenges 
the accepted contention that a transverse voltage cannot arise from a 
purely dissipative TDGL model.\cite{Kop,diss} Note however that the 
voltage 
we have found is statistically symmetric about zero, so that our result may be regarded as a sort of symmetry breaking.

It should be possible to use some of the techniques that are employed in visualization of vortices (such as STM) for the visualization of the islands predicted here.

\begin{acknowledgments}
This work has been supported in part by the Israel Science Foundation. I would like to acknowledge the IMA at the University of Minnesota for the hospitality during part of the time devoted to this study. I wish
to thank Gustavo Buscaglia for sending me the program used in Ref.~\onlinecite{Bolech}.
\end{acknowledgments}

\end{document}